\documentclass{aa}  

\usepackage{graphicx}
\usepackage{txfonts}
\usepackage[breaklinks=true,colorlinks=true,citecolor=blue]{hyperref} %% to avoid \citeads line fills
\begin{document} 

   \title{Source location and evolution of a multi-lane type II radio burst}

  % \subtitle{I. Overviewing the $\kappa$-mechanism}

   \author{P. Zucca
          \inst{1}\fnmsep\thanks{Corresponding Author \email{zucca@astron.nl}}
          \and
          P. Zhang\inst{2,7}
          \and
          K. Kozarev\inst{3}
          \and
          M. Nedal\inst{4}
          \and
          M. Mancini\inst{1}
          \and
          A. Kumari\inst{5}
          \and
          D. E. Morosan\inst{6,8}
          \and
          B. Dabrowski\inst{9}
          \and
          P. T. Gallagher\inst{4}
          \and
          A. Krankowski\inst{9}
          \and
          C. Vocks\inst{10}
          }

   \institute{ASTRON - Netherlands Institute for Radio Astronomy, Oude Hoogeveensedijk 4, 7991 PD Dwingeloo, The Netherlands 
         \and
             Center for Solar-Terrestrial Research, New Jersey Institute of Technology, Newark, NJ 07102, USA
         \and
            Institute of Astronomy and National Astronomical Observatory, Bulgarian Academy of Sciences, 72 Tsarigradsko Chaussee Blvd., 1784 Sofia, Bulgaria
        \and
        Astronomy \& Astrophysics Section, School of Cosmic Physics, Dublin Institute for Advanced Studies, DIAS Dunsink Observatory, Dublin D15 XR2R, Ireland
        \and 
        Udaipur Solar Observatory, Physical Research Laboratory, Dewali, Badi Road, Udaipur - 313001, Rajasthan, India
        \and
        Department of Physics and Astronomy, University of Turku, FI-20014 Turku, Finland
        \and
        Cooperative Programs for the Advancement of Earth System Science, University Corporation for Atmospheric Research, Boulder, CO, USA
        \and
        Turku Collegium for Science, Medicine and Technology, University of Turku, 20014, Turku, Finland
        \and
        Space Radio-Diagnostics Research Centre, University of Warmia and
Mazury, R. Prawochenskiego 9, 10-719 Olsztyn, Poland
        \and
        Leibniz-Institut für Astrophysik Potsdam (AIP)
An der Sternwarte 16, D-14482 Potsdam
        }

   \date{Received \today}

% \abstract{}{}{}{}{} 
% 5 {} token are mandatory
 
  \abstract
   {Shocks in the solar corona are capable of accelerating electrons that in turn generate radio emission known as type~II radio bursts. The characteristics and morphology of these radio bursts in the dynamic spectrum reflect the evolution of the shock itself, together with the properties of the local corona where the shock propagates.}
   {In this work, we study the evolution of a complex type~II radio burst showing a multi-lane structure, to find the locations where the radio emission is produced and relate them to the properties of the local environment where the shock propagates.}
   {Using radio imaging, we were able to track separately each lane composing the type II burst and relate the position of the emission to the properties of the ambient medium such as density, Alf\'ven speed, and magnetic field.}
   {We show that the radio burst morphology in the dynamic spectrum changes with time and it is related to the complexity of the local environment. The initial stage of the radio emission is characterized by a single broad lane in the spectrum, while the latter stages of the radio signature evolve in a multi-lane scenario. The radio imaging reveals how the initial stage of the radio emission separates with time into different locations along the shock front as the density and orientation of the magnetic field change along the shock propagation. At the time where the spectrum shows a multi-lane shape, we found a clear separation of the imaged radio sources propagating in regions with different density.}
  % conclusions heading (optional), 
   {By combining radio imaging with the properties of the local corona, we described the evolution of a type~II radio burst and, for the first time, identified three distinct radio emission regions above the CME front. Two regions were located at the flanks, producing earlier radio emission than the central position, in accordance with the complexity of density and Alfv\'en speed values in the regions where radio emission is generated. This unprecedented observation of a triple-source structure provides new insights into the nature of multi-lane type~II bursts.}


   \maketitle
%
%-------------------------------------------------------------------

\section{Introduction}\label{sec0}

% solar radio diagnose background structure
% solar radio in general
For more than half a century, solar radio burst observations have been an important tool used to diagnose the background plasma environment in the solar corona, including density distribution and magnetic field configuration \citep{wild1950typeII, Kontar2017NatCo, Chen2018NatAs}. These radio emission features can also provide information on the acceleration of fast electrons and identify the energy release processes and their locations in the solar corona \citep{2020FrASSsolaratmosphCostas,Dresing2022,morosan2025a}.
Solar radio bursts in the low-frequency radio range are categorized into 5 main types (Type~I-V) based on the dynamic spectrum morphology, each corresponding to specific physical processes occurring in the solar corona. Type~II radio bursts are generally accepted to be associated with shocks associated with coronal mass ejections (CMEs), \citep[e.g.,][]{mann2005electron,morosan2023type, ramesh2023ApJsolarcoronadens}, although a small but significant fraction can be produced by other mechanisms, like plasma jet eruptions \citep{Maguire2021lofar}, or flare-related blast waves and failed eruptions \citep{Magdalenic2012,Eselevich2013,Gopalswamy2016}.

%% multi-lane structures
Type~II radio bursts can exhibit complex features in their dynamic spectra, including band splitting and multiple emission lanes. These bursts typically show two main emission bands corresponding to fundamental (F) and harmonic (H) plasma emission, where the harmonic frequency is approximately twice the fundamental frequency \citep{Mann2005}. Band splitting, where either the fundamental or harmonic band appears to split into two closely spaced parallel lanes, is commonly observed \citep{Vrsnak2001,Vrsnak2002}. This splitting phenomenon has been traditionally interpreted as emission from the upstream and downstream regions of the shock \citep{Smerd1974,Smerd1975} and the observed emission is expected to be co-spatial in radio images. However, recent imaging studies of type II lanes have also identified type II sources at separate and distinct locations at the shock \citep{bhunia2022imagingspectroscopy, morosan2023type}. 

Multi-lane structures may result from different physical processes: (1) wave mode splitting between the plasma frequency ($\omega_p$) and the upper-hybrid frequency ($\sqrt{\omega_p^2 + \omega_g^2}$) \citep{Sturrock1961}, (2) simultaneous emission from different parts of the shock front with varying plasma densities \citep{zucca2014formation,Morosan2019a,bhunia2022imagingspectroscopy}, or (3) multiple shock waves propagating through the corona \citep{Eselevich2013,Zimovets2015AdSpR}. Despite these theoretical interpretations, direct imaging observations of multi-lane structures below 100~MHz have been limited until now, especially studies with imaging of the entire extent of type~II radio bursts in the decameter wavelengths. 

% recent progress on solar radio imaging
Recent progress in radio instrumentation is pushing forward the observation capability to resolve more and more details of the complex type II structures observed.
The LOw Frequency ARray (LOFAR) \citep{van2013lofar} plays a pivotal role in this evolution. LOFAR operates in the low-frequency range of 10 to 240~MHz. It has two sets of antennae: the Low band antenna (LBA) in 10-88\,MHz and the high band antenna (HBA) in 110-240~MHz. LOFAR's unique design, featuring a vast array of antenna stations spread over Europe, allows for high-resolution imaging and precise measurements.

% No dedicated interferometry is done
In this work, we present an observation of a type~II solar radio burst that initially shows a single emission lane and evolves into a complex multi-lane structure. For the first time in low-frequency radio observations, we identify three distinct emission regions along the CME-driven shock front, providing new insights into the spatial distribution of electron acceleration regions. This unique observation allows us to track the evolution of these separate emission sources and their relationship to the local coronal conditions.
% this paper 
This paper is presented as follows, Section \ref{sec1} presents the observation detail, including data processing and event overview. In Section \ref{sec2}, we demonstrate the analysis of the radio source location and relation to the corona. Finally, Section \ref{sec3} presents the discussion and conclusions.

\section{Observation}\label{sec1}
On 2022 May 19, a complex type~II radio burst was observed following a CME eruption. Figure 1 shows the composite dynamic spectrum combining observations from three instruments: the Observation Radio Fréquences Etalées of Spectrographie Solaire (ORFEES; \citep{Hamini2021}) operating in the frequency range 300-800~MHz, and the LOFAR High Band Antenna (HBA, 110-240~MHz) and Low Band Antenna (LBA, 30-88\,MHz) \citep{van2013lofar}. The LOFAR spectrum is pre-processed with an RFI-flagging tool for solar and spaceweather spectrum: \texttt{ConvRFI} \footnote{ConvRFI \href{https://github.com/peijin94/ConvRFI}{https://github.com/peijin94/ConvRFI}} \citep{zhang2023MNRASConvRFI}. 

The type~II burst started at 12:02 UT with emission observed at frequencies around 600~MHz, drifting to lower frequencies over time. At high frequencies (>200~MHz), the fundamental (F) and harmonic (H) emission bands appear superposed, while they become clearly separated in the LOFAR LBA frequency range. The event initially shows a single/double lane structure at higher frequencies, but evolves into a complex multi-lane pattern particularly evident in the LOFAR-LBA range (30-88\,MHz) after 12:08 UT.

Similarly, as the spectrum also the radio imaging show an evolution of the complexity of the radio source location with time. Initially, a single radio source is identified along the shock front. Figure 2 presents the Solar Dynamics Observatory (SDO) Atmospheric Imaging Assembly (AIA; \citet{Lemen2012}) running difference images at 193~\AA, showing the CME eruption, with superposed Nançay Radioheliograph (NRH; \citep{Kerdraon1997}) radio contours at 432~MHz. The NRH data were processed using the standard SolarSoft (SSW) IDL package available in the SSW/radio/nrh directory, which includes routines for amplitude and phase calibration using the quiet Sun as reference \citep{mercier2015}. The NRH observations at high frequencies show a single radio source location associated with the type~II and the erupting CME front.

\begin{figure}
    \centering
    \includegraphics[width=0.49\textwidth]{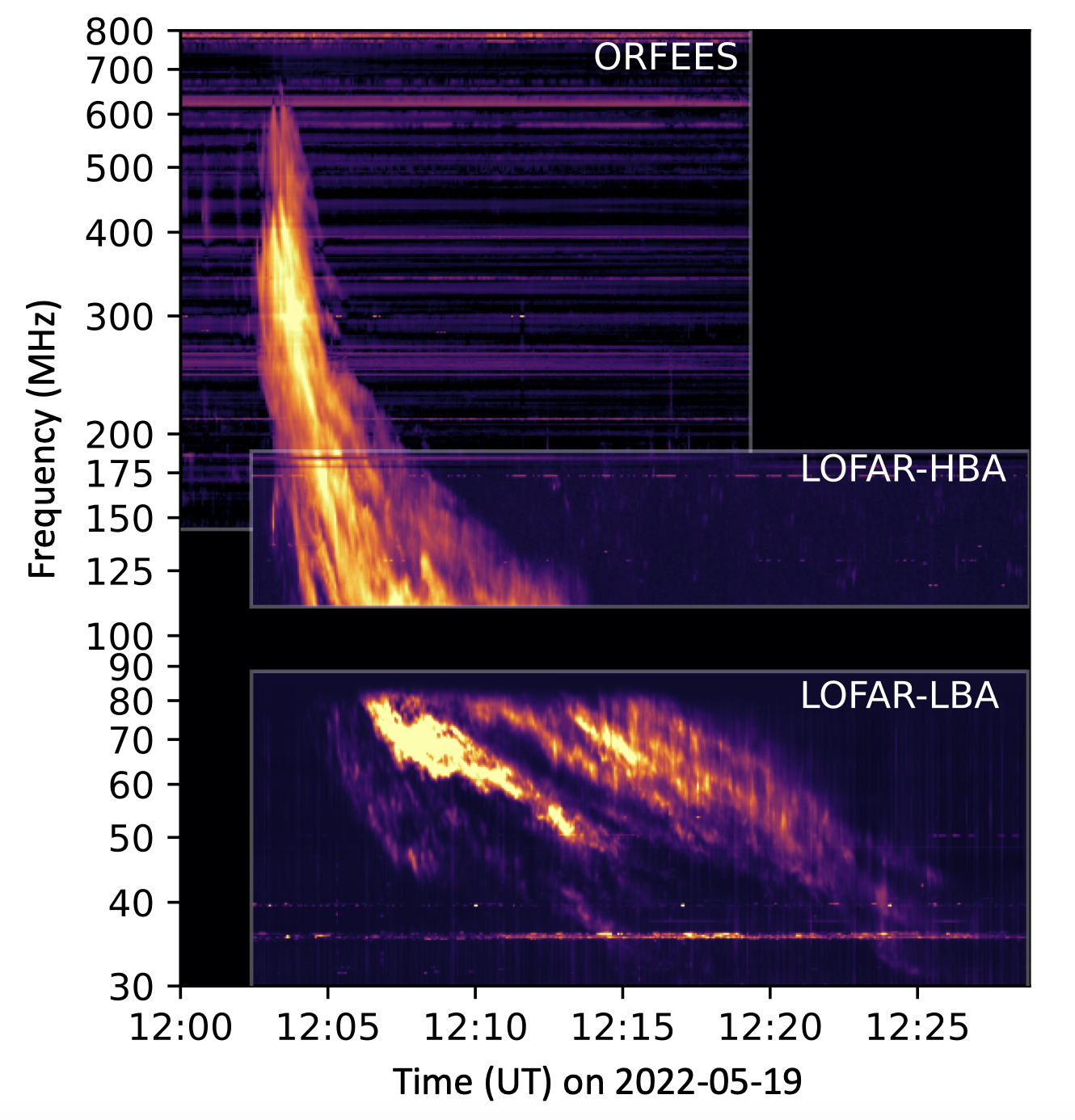}
    \caption{Composite dynamic spectrum showing the evolution of the type~II radio burst observed on 2022-May-19. The spectrum combines observations from three instruments: ORFEES (300-800~MHz), LOFAR High Band Antenna (HBA, 110-240 MHz), and LOFAR Low Band Antenna (LBA, 30-88\,MHz). The event starts at 12:02~UT with a single emission lane at frequencies around 600 MHz and drifts to lower frequencies. The multi-lane structure becomes particularly evident in the LOFAR-LBA frequency range (30-88\,MHz) after 12:08~UT. Note that fundamental (F) and harmonic (H) emissions are superposed at high frequencies, in the LBA range they are distinct. The frequency range is plotted on a logarithmic scale to better display the fine structures across the wide frequency range.}
    \label{fig:ds_all}
\end{figure}

\begin{figure}
    \centering
    \includegraphics[width=0.49\textwidth]{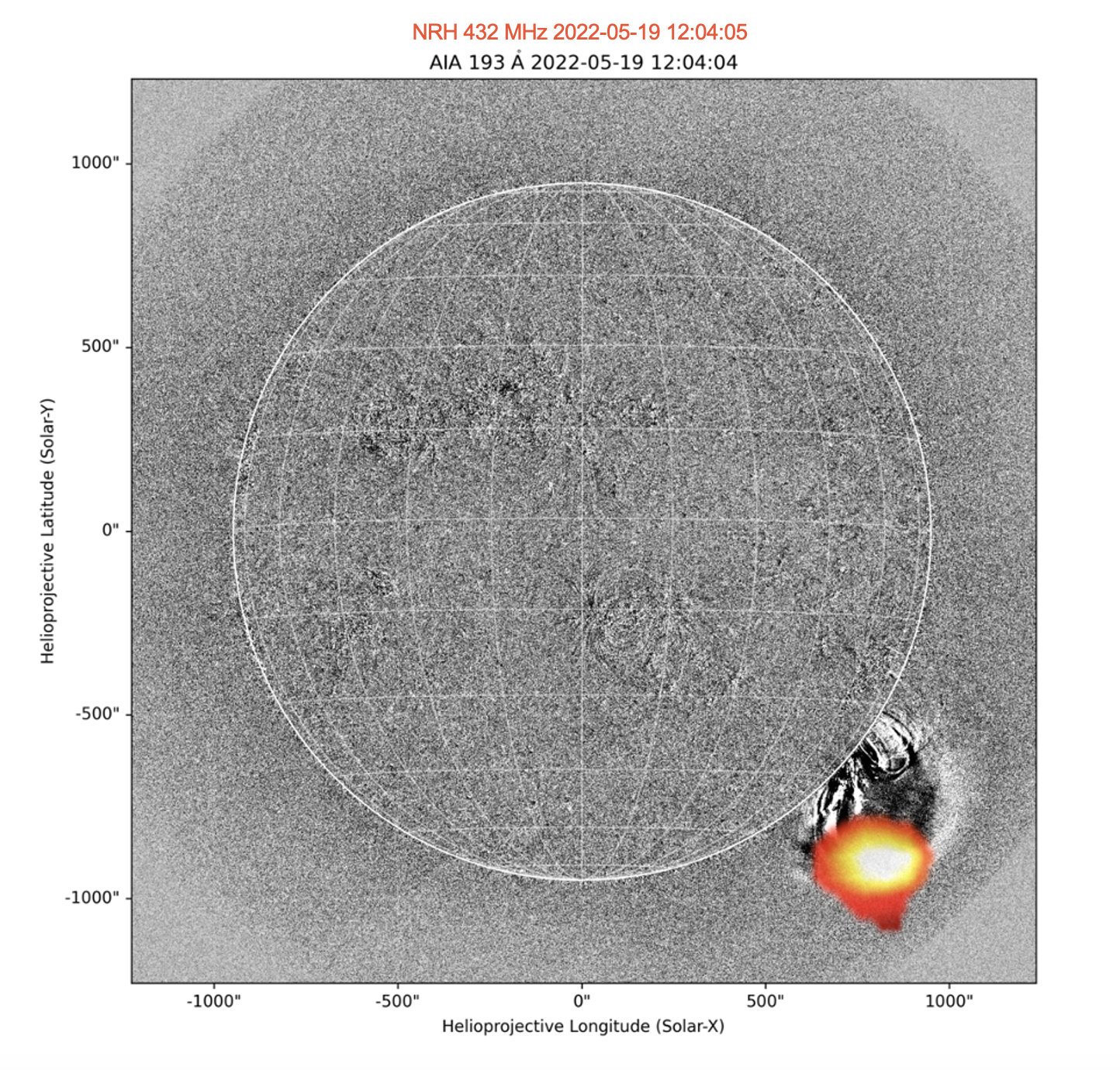}
    \caption{Composite image showing the CME eruption observed on 2022-May-19. The background shows the SDO/AIA running difference image at 193~$\AA$ revealing the CME front structure. The radio contours (in yellow/red) from the Nançay Radioheliograph (NRH) at 432~MHz show a single radio source location associated with the type~II radio burst and the erupting CME front at 12:04~UT. The radio source at high frequencies appears as a single emission region, in contrast to the multiple emission regions observed at lower frequencies with LOFAR.}
    \label{fig:NRH}
\end{figure}

\subsection{Imaging the type II multilanes with LOFAR}

In this section, we present the observation of the spatial distribution of the various emission lanes observed with LOFAR LBA interferometric imaging. The LBA frequency range (30-88\,MHz) is particularly suitable for this analysis as the fundamental (F) and harmonic (H) emission bands are clearly separated, allowing us to independently study their source locations and evolution. The availability of interferometric imaging at these frequencies enables us to track the position and development of each emission lane throughout the event.
The imaging in this work is performed with a combination of Core and Remote Stations of LOFAR.
Radio imaging allows for arcmin-scale spatial resolution in decameter wavelength \citep{Zhang2022lofarsun,morosan2025}. 
The LOFAR interferometry data processing consists of several sequential steps. First, we perform gain calibration by applying DP3 (Default Pre-Processing Pipeline) to the Cas-A calibrator data to compute phase and amplitude offsets, using a flux density model as a reference. This generates gain solutions for each baseline and subband. DP3 (The Default Pre-Processing Pipeline)\footnote{DP3 \href{https://github.com/lofar-astron/DP3}{https://github.com/lofar-astron/DP3}} \citep{2018asclsoft04003V}. Following this, we conduct antenna inspection by examining phase and amplitude plots for each antenna to identify and flag corrupt data. The next step involves applying calibration corrections to phase and amplitude in the target observation using the gain solutions from DP3. For imaging, we transform visibility data from wave vector space $[u,v]$ to image space $[x,y]$ using a 2D Fourier Transform, employing the w-stacking CLEAN algorithm with wsclean for point-spread-function (PSF) deconvolution\footnote{wsclean \href{https://gitlab.com/aroffringa/wsclean}{https://gitlab.com/aroffringa/wsclean}} \citep{offringaWsclean2014}. Finally, in post-processing, we transform coordinates to helio-projective and convert units to Brightness temperature for further analysis using lofarSun\footnote{lofarSun \href{https://github.com/peijin94/LOFAR-Sun-tools}{https://github.com/peijin94/LOFAR-Sun-tools}} \citep{Zhang2022lofarsun}.

\begin{figure*}
    \centering
    \includegraphics[width=0.93\textwidth]{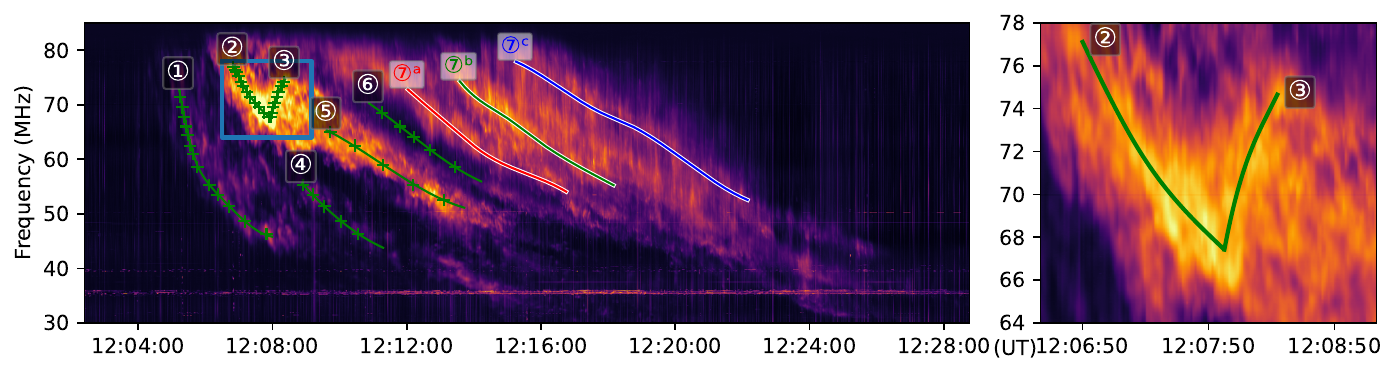}
    \includegraphics[height=0.26\textwidth]{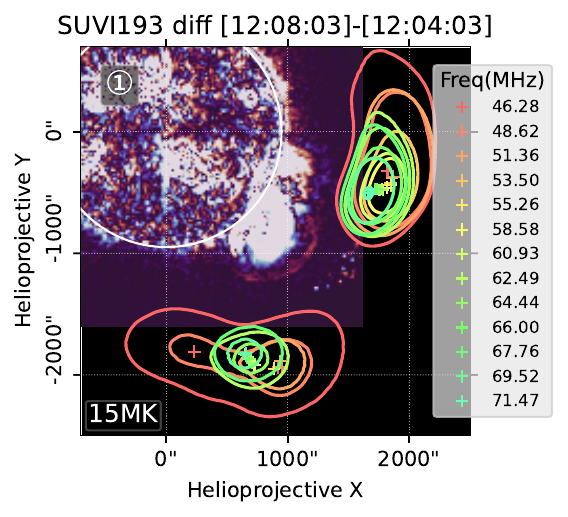}
    \includegraphics[height=0.26\textwidth]{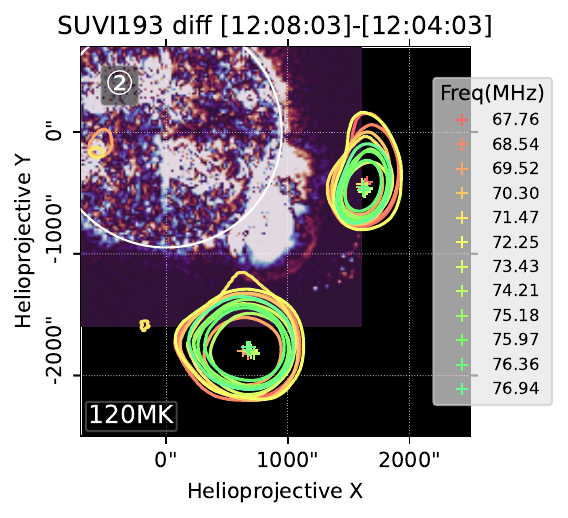}
    \includegraphics[height=0.26\textwidth]{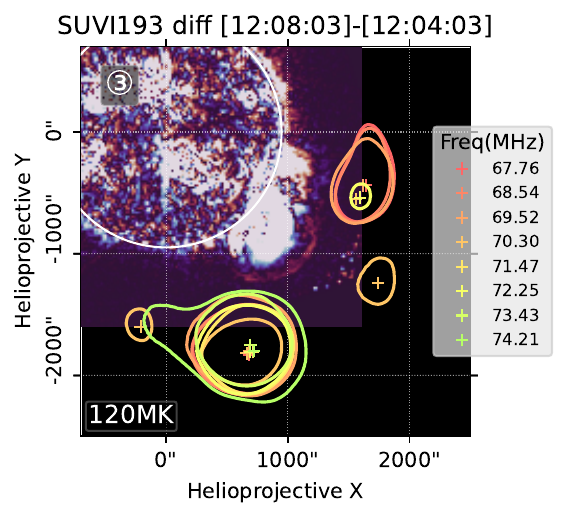}
    \includegraphics[trim={0cm 0.8cm
    0 0.2cm},clip,width=0.93\textwidth]{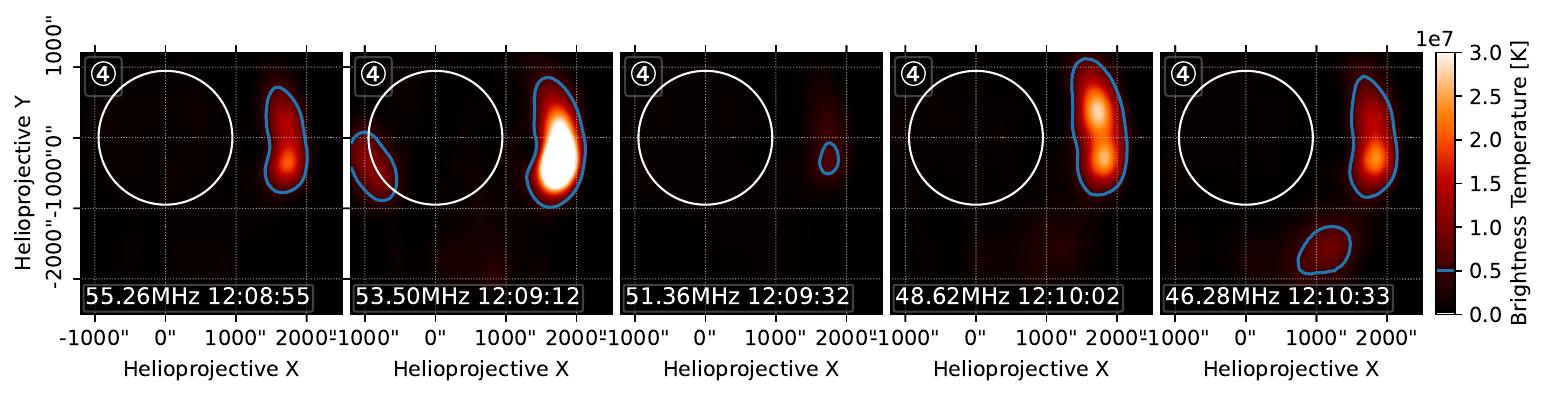}
     \includegraphics[trim={0cm 0.8cm
    0 0.2cm},clip,width=0.93\textwidth]{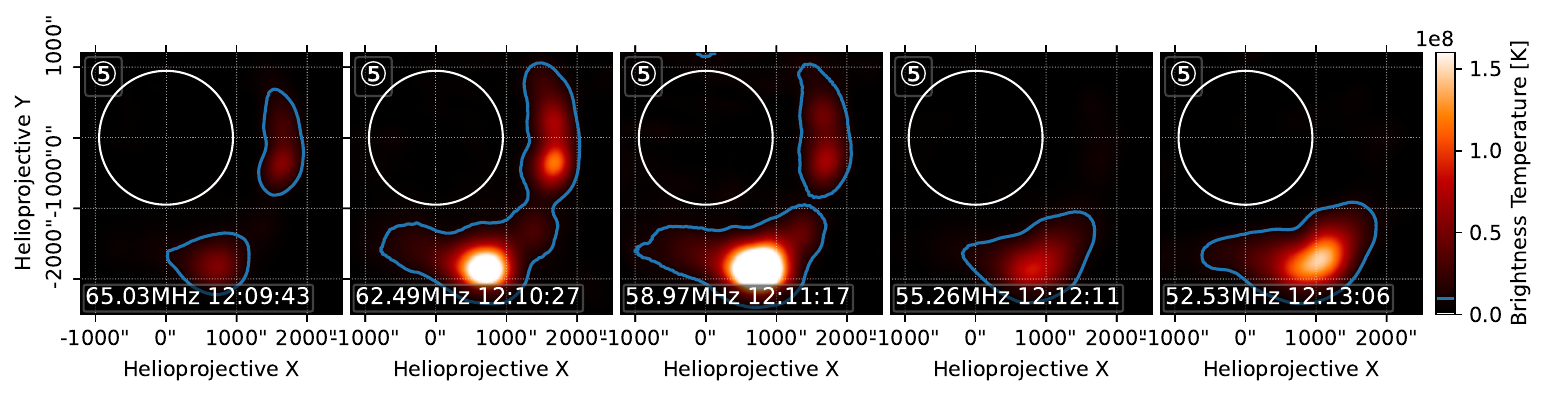}
         \includegraphics[trim={0cm 0.8cm
    0 0.2cm},clip,width=0.93\textwidth]{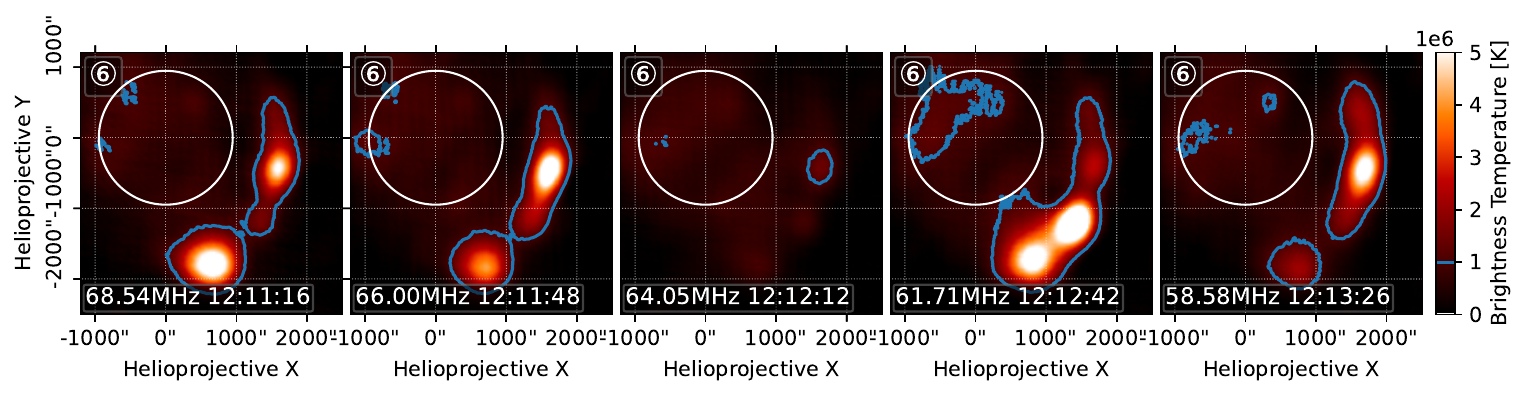}
    \caption{Overview of the type~II radio burst lanes observed with LOFAR-LBA. Top panel: Dynamic spectrum showing the frequency range 30-88\,MHz, with marked regions indicating different emission features. Regions 1-5 correspond to fundamental plasma emission: lanes 1-2 show the initial double source structure, while region 3 marks the appearance of a third distinct emission source. Lanes 4-5 continue to show fundamental emission with varying spatial distributions. Lane 6 shows emission comparable to the quiet Sun level. Regions 7a-c, appearing later in time, represent harmonic emission showing three distinct parallel lanes. Lower panels: LOFAR radio imaging of lanes 1-6, showing the spatial distribution of each emission feature. Lanes 1-3 show similar source locations with double source structures on the eastern and southern sides of the shock front. Lane 4 shows emission predominantly on the eastern side, while lane 5 exhibits complex structures in both eastern and southern regions. The detailed imaging of the three parallel harmonic lanes (7a-c) is presented in Figures 4 and 5.}
    \label{fig:dsim_main}
\end{figure*}

To describe the observation of this complex type~II event, we imaged each lane of the type II burst identified in the dynamic spectrum in the LOFAR-LBA frequency range (30-88\,MHz). Figure (\ref{fig:dsim_main}) shows an overview of the lanes observed with LOFAR-LBA. The top panel presents the dynamic spectrum with seven selected lanes (including a quiet period), corresponding to different emission features observed during the event. The fundamental plasma emission is represented by the type II lanes labelled 1--5, while region 6 corresponds to a period of quiet emission, and labels 7a--c show the harmonic emission appearing later in the event as three distinct parallel lanes. 

The top right inset of Figure 3 shows a zoomed view of the dynamic spectrum during an unusual phase of the event, highlighting region 3 where a third emission source begins to appear in both the spectrum and the corresponding radio imaging. This marks a transition from the initial double-source structure to a more complex configuration with three distinct emission regions, the imaging of this transition is shown in panels marked with 2 and 3 in Figure 3.

The lower panels of Figure 3 present the LOFAR interferometric images corresponding to regions 1-6. The fundamental emission lanes (1-5) show an evolution in their spatial distribution: lanes 1-3 exhibit similar source locations with double source structures on the eastern and southern sides of the shock front, lane 4 shows emission predominantly on the eastern side, while lane 5 displays complex structures in both eastern and southern regions. Lane 6, corresponding to a period of reduced emission in the spectrum, shows brightness temperature levels comparable to the quiet Sun. The three parallel harmonic lanes (7a-c), appearing later in the event, are presented in detail in Figures 4 and 5, where their distinct spatial separation perpendicular to the shock propagation direction becomes evident.

\begin{figure}
    \centering
    \includegraphics[width=\columnwidth]{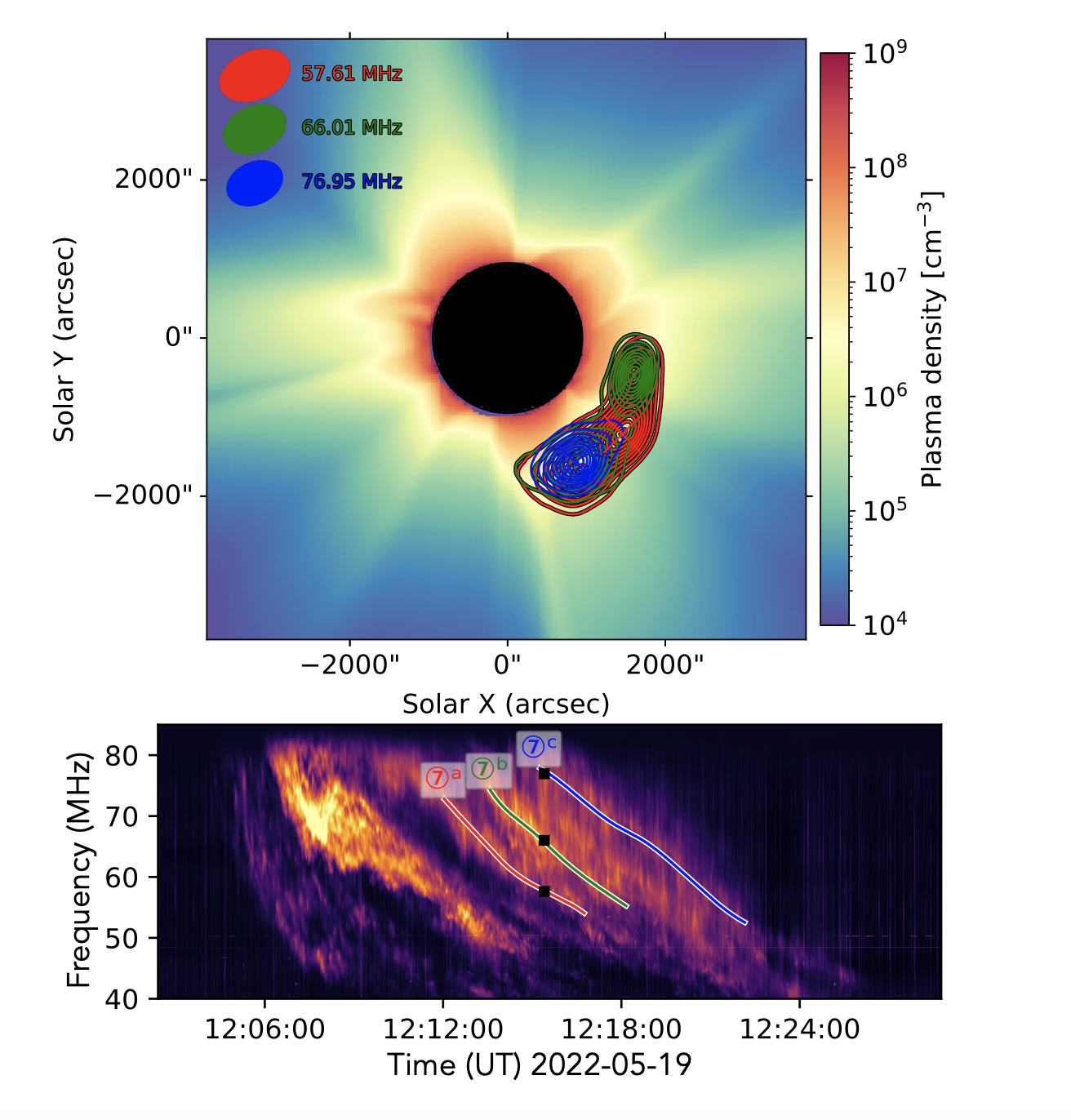}
    \caption{MAS MHD model results. Density and three lanes in spectra.}
    \label{fig:MAS_PFSS}
\end{figure}

\begin{figure}
    \centering
    \includegraphics[width=\columnwidth]{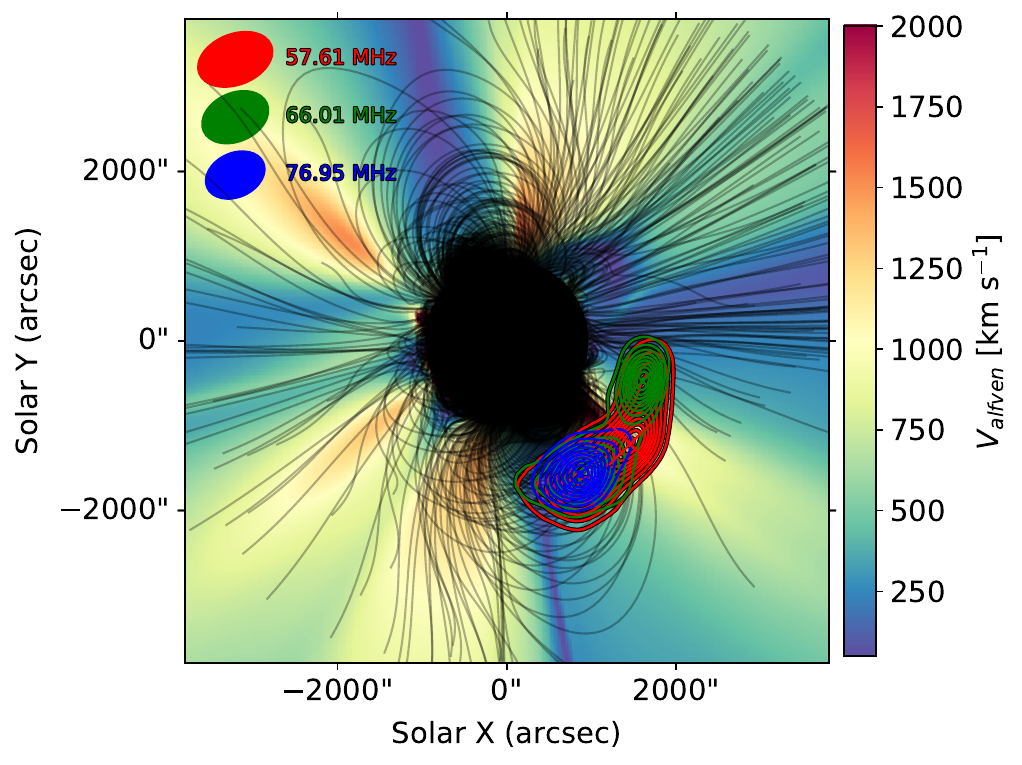}
    \caption{MAS MHD model results. panel shows Alfv\'en speed with the PFSS model and LOFAR contours on top of it.}
    \label{fig:MAS_PFSS}
\end{figure}

\section{Results}\label{sec2}
The imaging analysis of this type~II radio burst reveals several key features that evolve with time. Initially, at high frequencies (>200 MHz), we observe a single radio source associated with the CME front, as shown by the NRH observations (Figure 2). However, as the event progresses to lower frequencies, the radio emission becomes increasingly complex, revealing multiple distinct source regions.

In the LOFAR-LBA frequency range (30--88\,MHz), we identify seven distinct emission features (Figure 3). Lanes 1-3 show fundamental plasma emission with similar source locations, characterized by double source structures positioned on the eastern and southern sides of the shock front, above the EUV front observed in SUVI imagery. During this phase, we observe the emergence of a third distinct emission source. Lane 4 exhibits fundamental emission predominantly localized to the eastern side of the shock front, while Lane 5 shows more complex fundamental emission structures distributed across both eastern and southern regions. Lane 6 represents a transition period where the emission intensity becomes comparable to the quiet Sun level, displaying complex spatial structures. The most notable feature appears in region 7, which splits into three parallel lanes (7a-c) in the harmonic emission band. These lanes show clear spatial separation perpendicular to the shock propagation direction (radial), a feature not previously observed in type~II bursts at these frequencies.

The comparison between the observed radio source locations and coronal parameters derived from the Predictive Science Magnetohydrodynamics Algorithm outside a Sphere (MAS) model \citep{Riley2011} provides additional insights into the physical conditions governing the multi-lane emission. The three parallel lanes (7a-c) show a clear correspondence between their frequency drift rates and the local plasma densities in their respective source regions. As shown in Figure 4, the red source appears in a region of lower density, corresponding to the lower-frequency lane in the dynamic spectrum, while the blue source originates from a region of higher density, consistent with its higher-frequency emission. The green source, appearing in a region of intermediate density, produces the middle-frequency lane. This spatial-spectral correlation demonstrates how the multi-lane structure in the dynamic spectrum directly reflects the density stratification along different parts of the shock front.

Figure 5 presents the Alfv\'en speed distribution derived from the same MHD model, revealing enhanced Alfv\'en speeds in the region corresponding to the central (red) source. This higher Alfv\'en speed explains the delayed appearance of the central lane compared to the flank emissions, as shock formation and particle acceleration are less efficient in regions of higher Alfv\'en speed. The two flank regions (green and blue sources) experience lower Alfv\'en speeds, facilitating earlier shock formation and radio emission.

\section{Conclusion and discussion}\label{sec3}

We present a comprehensive observational study of a CME-driven shock and its associated type~II radio emission, revealing an unprecedented evolution from a single radio source to a complex multi-lane structure. The dynamic spectrum shows increasing complexity over time, mirrored in the spatial distribution of radio sources, which evolves from a single emission region to three distinct sources along the shock front.

Our analysis demonstrates a clear correlation between the spectral characteristics of individual lanes and the local coronal conditions where they originate. The frequency drift rates of individual lanes show strong agreement with the plasma densities in their respective source regions, as derived from MHD modeling using the Predictive Science MAS model \citep{Riley2011}. Furthermore, we observe a temporal correspondence between the appearance of radio sources and the local Alfv\'en speed distribution along the shock front, with three distinct emission regions developing along different parts of the CME-driven shock.

Previous studies \citep[e.g.,]{zucca2014b,carley2013,kong2012,Zucca2018AAShock} have shown that CME flanks typically provide favorable conditions for radio emission and particle acceleration. Our observations extend this understanding by showing that multiple lanes in type~II bursts can arise from distinct acceleration regions, each characterized by specific Mach numbers and density profiles along the shock front.

While our findings do not invalidate other mechanisms for band-splitting or multiple lanes in type~II bursts, such as upstream/downstream shock emission \citep{Vrsnak2001,Vrsnak2002}, they provide compelling evidence that in this event, the multi-lane structure directly reflects different acceleration regions along the CME shock front. The spatial-spectral correspondence we observe between lane frequencies and source locations, and between source activation times and local Alfv\'en speeds, strongly supports this interpretation.

These observations highlight the crucial role of radio spectro-heliographic imaging, as provided by LOFAR, in understanding complex solar radio phenomena. Further studies of both multi-lane events and simpler single-lane or band-split type~II bursts are essential to fully comprehend the variety of physical processes involved in coronal shock acceleration and radio emission. Such investigations will benefit from the continued development of high-resolution, low-frequency imaging capabilities \citep{Morosan2019a,Zhang2022lofarsun}.

\begin{acknowledgements}
This paper is based on data obtained with the International LOFAR Telescope (ILT) under project code LT16-001 with PI Dr. P. Zucca. LOFAR (van Haarlem et al. 2013) is the Low Frequency Array designed and constructed by ASTRON. It has observing, data processing, and data storage facilities in several countries, that are owned by various parties (each with their own funding sources), and that are collectively operated by the ILT foundation under a joint scientific policy. The ILT resources have benefited from the following recent major funding sources: CNRS-INSU, Observatoire de Paris and Université d'Orléans, France; BMBF, MIWF-NRW, MPG, Germany; Science Foundation Ireland (SFI), Department of Business, Enterprise and Innovation (DBEI), Ireland; NWO, The Netherlands; The Science and Technology Facilities Council, UK; Polish Ministry of Science and Higher Education number: 2021/WK/2. KK acknowledges support by the Bulgarian National Science Fund, VIHREN program, under contract KP-06-DV-8/18.12.2019 (MOSAIICS project). D.E.M. acknowledges the Research Council of Finland project `SolShocks' (grant number 354409). MN acknowledges support by the project "The Origin and Evolution of Solar Energetic Particles”, funded by the European Office of Aerospace Research and Development under award No. FA8655-24-1-7392.

SDO data are courtesy of NASA/SDO and the AIA science teams. SUVI data are courtesy of NOAA/NESDIS/NCEI. The authors thank the Predictive Science team (https://www.predsci.com/) for making their MHD simulation results available.

The authors would like to thank the radio observatory of Nançay for making the NRH and ORFEES data available.

This research used the SunPy open source software package \citep{Sunpy:2020}.
      
\end{acknowledgements}

\bibliographystyle{aa} % style aa.bst
\bibliography{cite} % your references Yourfile.bib

\end{document}